\documentclass[sn-mathphys,Numbered]{sn-jnl}

\usepackage{graphicx}%
\usepackage{multirow}%
\usepackage{amsmath,amssymb,amsfonts}%
\usepackage{amsthm}%
\usepackage{mathrsfs}%
\usepackage[title]{appendix}%
\usepackage{xcolor}%
\usepackage{textcomp}%
\usepackage{manyfoot}%
\usepackage{booktabs}%
\usepackage{algorithm}%
\usepackage{algorithmicx}%
\usepackage{algpseudocode}%
\usepackage{listings}%
\usepackage{physics} 

\theoremstyle{thmstyleone}%
\theoremstyle{thmstyletwo}%

\theoremstyle{thmstylethree}%

\begin{document}

\title[Extrapolative machine learning in materials research]{Advancing Extrapolative Predictions of Material Properties through Learning to Learn}


\author[1]{\fnm{Noda} \sur{Kohei}}
\email{Kouhei\_Noda@jsr.co.jp}

\author[1]{\fnm{Wakiuchi} \sur{Araki}}

\author[2,3]{\fnm{Hayashi} \sur{Yoshihiro}}

\author[2,3]{\fnm{Yoshida} \sur{Ryo}}\email{yoshidar@ism.ac.jp}

\affil[1]{\orgdiv{RD Technology and Digital Transformation Center}, \orgname{JSR Corporation}, \orgaddress{\city{Yokkaichi}, \postcode{510-8552}, \country{Japan}}}

\affil[2]{\orgdiv{The Institute of Statistical Mathematics}, \orgname{Research Organization of Information and Systems}, \orgaddress{\city{Tachikawa}, \postcode{190-8562}, \country{Japan}}}

\affil[3]{\orgdiv{The Graduate Institute for Advanced Studies}, \orgname{SOKENDAI}, \orgaddress{\city{Tachikawa}, \postcode{190-8562}, \country{Japan}}}



\abstract{
Recent advancements in machine learning have showcased its potential to significantly accelerate the discovery of new materials. Central to this progress is the development of rapidly computable property predictors, enabling the identification of novel materials with desired properties from vast material spaces. However, the limited availability of data resources poses a significant challenge in data-driven materials research, particularly hindering the exploration of innovative materials beyond the boundaries of existing data. While machine learning predictors are inherently interpolative, establishing a general methodology to create an extrapolative predictor remains a fundamental challenge, limiting the search for innovative materials beyond existing data boundaries.  In this study, we leverage an attention-based architecture of neural networks and meta-learning algorithms to acquire extrapolative generalization capability. The meta-learners, experienced repeatedly with arbitrarily generated extrapolative tasks, can acquire outstanding generalization capability in unexplored material spaces. Through the tasks of predicting the physical properties of polymeric materials and hybrid organic--inorganic perovskites, we highlight the potential of such extrapolatively trained models, particularly with their ability to rapidly adapt to unseen material domains in transfer learning scenarios.}


\maketitle

\section*{Introduction}\label{sec1}

In recent years, the potential of machine learning to accelerate the process of discovering new materials has been demonstrated across diverse material systems, such as polymers \citep{Wu2019-oz}, inorganic compounds \citep{Merchant2023-yr, szymanski2023autonomous}, alloys \citep{Rao2022-bq}, catalysts \citep{Zhong2020-ud, Kim2020-ze}, aperiodic materials \citep{Liu2021-ft, liu2023quasicrystals, uryu2024deep}. At the heart of this advancement lies a rapidly computable property predictor obtained through machine learning that represents the compositional and structural features of any given material in a vector form and learns the mathematical mapping from such vectorized materials to their physicochemical properties. By employing such a property predictor with millions or even billions of candidate materials, novel materials with tailored properties can be identified effectively by navigating the expansive search space. 

The most significant challenge in such data-driven materials research is the scarcity of data resources \citep{Coley2020-in, Martin2023-xd, Tu2023-xl}. In most research tasks, ensuring sufficient quantity and diversity of data remains a formidable hurdle. Furthermore, the ultimate goal of materials science is to discover ``innovative'' materials that exist in a material space no one has gone before. However, machine learning is generally interpolative, and its predictability is limited to the domain neighboring the given training data. Even large language models, currently revolutionizing various fields, are essentially memorization learners, making interpolative predictions based on vast data. Establishing fundamental methodologies for extrapolative predictions poses an unsolved challenge not only in materials science but also in the next generation of artificial intelligence \citep{Meredig2018-yx, Xiong2020-in, Shimakawa2024-cf}.

Methodological research related to extrapolative machine learning has progressed within various frameworks, including domain generalization \citep{Zhou2022-zr, Wang2022-ak}, transfer learning \citep{Pan2010-yt}, domain adaptation \citep{Wilson2020-gb, Farahani2021-yk}, meta-learning \citep{Hospedales2022-lv}, and multi-task learning \citep{Caruana1997-ib}, all of which are closely interrelated. These methodologies seek to overcome the challenge of limited data availability by integrating heterogeneous datasets with different generative processes, from the source and target domains. Wu et al. (2019) employed transfer learning to successfully discover three new amorphous polymers with notably high thermal conductivity \citep{Wu2019-oz}. Given the limited availability of thermal conductivity data for only 28 amorphous polymers in the target domain, they constructed a transferred model for thermal conductivity prediction by refining a collection of source models, pre-trained on other related properties, such as glass transition temperature, specific heat, and viscosity, for which a well-supplied dataset existed. Remarkably, the dataset for the target task lacked similar instances for the three synthesized polymers. Nevertheless, the transferred model exhibited out-of-distribution generalization performance, attributed to the presence of relevant cases in the source datasets. In materials research, several instances have been reported where transfer learning successfully acquired extrapolative capabilities \citep{Yamada2019-sz, Ju2021-qn}. In the growing fields of artificial intelligence, such as computer vision and natural language processing, research on domain generalization is much more active than in materials science \citep{Zhou2022-zr, Wang2022-ak}. In domain generalization, for example, numerous sets of data from different domains, called episodes, are generated from the entire given dataset, and the model undergoes domain adaptation repeatedly \citep{Li2018-sa, Balaji2018-cj}. During this repeated training, the resulting model often acquires domain-invariant feature representations, thus achieving generalization performance for unseen domains. For example, a set of episodes can be generated by manipulating an original image with varying appearances, brightness, and backgrounds. In materials research, different material classes, such as polyester or cellulose, could correspond to different domains. However, it remains uncertain whether the generic methodologies of domain generalization can maintain effectiveness in materials property prediction tasks. It is intuitively plausible that there exists a domain-invariant representor or predictor across synthetically manipulated images. However, it is not obvious that such invariance exists in different material systems.

In this study, we leverage an attention-based architecture originally designed for few-shot learning, referred to as matching neural networks (MNNs) \citep{Vinyals2016-nl}, to learn the learning method for obtaining extrapolative predictors. We employ the meta-learning algorithm \citep{Koch2015-el, Vinyals2016-nl, Ravi2017-fl,Snell2017-yl, Finn2017-ae}, commonly known as ``learning to learn'', to achieve extrapolative prediction capability and out-of-distribution generalization performance. From a given dataset $\mathcal{D}$, numerous episodes are generated, each comprising a training set $\mathcal{S}$ and a test set $\mathcal{Q}$ containing instances outside the training domain $\mathcal{S}$. The objective is to learn a generic model $y=f(x, \mathcal{S})$ representing the mapping from material $x$ to property $y$ in which $(x, y)$ belongs to any domain $\mathcal{Q}$. A distinctive feature of MNNs is to explicitly include the training dataset $\mathcal{S}$ as an input variable. Instances of the input-output pair $(x, y)$ are assumed to follow a distribution different from $\mathcal{S}$. Unlike other domain adaptation methods, MNNs explicitly describe in the model $y=f(x, \mathcal{S})$ how it predicts $y$ from $x$ in an unseen domain given a training dataset $\mathcal{S}$. 

In the following, we demonstrate how the extrapolatively trained predictors acquire extrapolation capabilities through two property prediction tasks for polymeric materials and hybrid organic--inorganic perovskite compounds. For a given dataset, we can generate a set of episodes for extrapolative learning, flexibly in terms of quantity and quality. This is considered a form of self-supervised learning. As shown later, the condition of generating episode sets, such as the overall data size and the size of $\mathcal{S}$ in the training and inference phases, significantly influences the resulting generalization performances. Through various numerical experiments, we provide guidelines for configuring these parameters. Moreover, we use the extrapolatively trained predictor as a pre-trained model for downstream tasks, adapting it to the target domain using data from an extrapolative domain of the material space. The extrapolatively trained predictor exhibits remarkable transferability, adapting to the downstream extrapolative prediction tasks with much smaller training instances, compared to conventionally trained models.

\section*{Results}\label{sec2}

\subsection*{Methods outline}

\begin{figure}[h]%
    \centering
    \includegraphics[width=\textwidth]{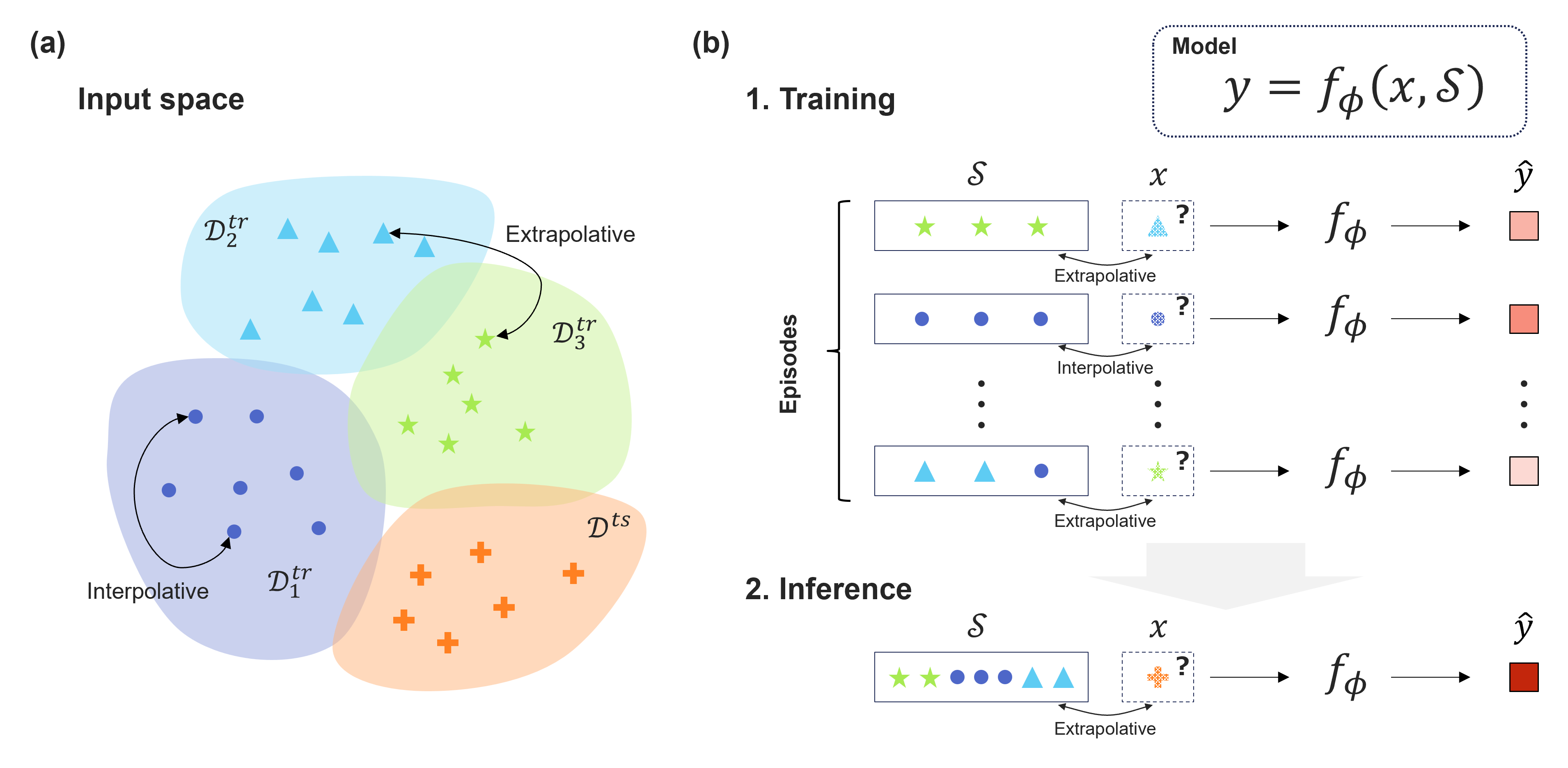}
    \caption{Extrapolative episodic training (E2T) with MNNs involves generating numerous episodes from a given dataset, comprising a support set ($\mathcal{S}$) and an input-output pair $(x,y)$. By including a large number of $\mathcal{S}$ and $(x,y)$ with extrapolative relationships into the episode set, the trained MNN learns the general way $y = f(x, \mathcal{S})$ for predicting extrapolatively from $x$ to $y$ with any given $\mathcal{S}$.}\label{fig:E2T_overview}
\end{figure}

A conventional machine learning predictor describes the relationship between input $x$ and output $y$ as $y = f_{\phi}(x)$. After training the model, the parameter $\phi$ is given as an implicit functional of the training dataset $\mathcal{S}$ as $y = f_{\phi(\mathcal{S})}(x)$. In contrast, the meta-learner $y = f_{\phi}(x, \mathcal{S})$ takes both the input--output variables $(x, y)$ and the training dataset $\mathcal{S} = \{x_i, y_i | i=1,\ldots, m\}$ consisting of $m$ instances, as its arguments. In the context of meta-learning, $\mathcal{S}$ is referred to as the support set. We will use this term hereafter. From a given dataset $\mathcal{D} = \{x_i, y_i | i=1, \ldots, d\}$, a collection of $n$ training instances, referred to as episodes, is constructed as $\mathcal{T} = \{x_i, y_i, \mathcal{S}_i | i= 1,\ldots, n\}$ to train the meta-learner. In this scenario, for each episode $(x_i, y_i)$ and $\mathcal{S}_i$, tuples in an extrapolative relationship can be arbitrarily chosen. For instance, $(x_i, y_i)$ represents a physical property $y_i$ of a cellulose derivative $x_i$, while $\mathcal{S}_i$ represents a dataset from other polymer classes, such as conventional plastic resins. Alternatively, $(x_i, y_i)$ can be defined by a compound containing element species that are not present in the training compounds comprising $\mathcal{S}_i$. An essential aspect here is that such extrapolative episodes can be arbitrarily generated from a given dataset. We refer to such a learning scheme as extrapolative episodic training (E2T) (Fig. \ref{fig:E2T_overview}).

This study focuses on real-valued output $y \in \mathbb{R}$ representing a physical property. Our model is based on an attention-based neural network that associates input and output variables as follows:
\begin{eqnarray}
y = \sum_{(x_i, y_i) \in \mathcal{S}} a(\phi_x, \phi_{x_i}) y_i = \vb{a}(\phi_x)^{\top} \vb{y}
\label{eq:attention}
\end{eqnarray}
Here the output $y$ is computed by taking the weighted sum of $y_i$ within the support set $\mathcal{S}$ using the weight $a(\phi_x, \phi_{x_i})$. The second equation represents this in a vector form with $ \vb{y}^{\top} = (y_1, \ldots, y_m) \in \mathbb{R}^m$ and $\vb{a}(x)^{\top} = ( a(\phi_x, \phi_{x_1}), \ldots, a(\phi_x, \phi_{x_m})) \in \mathbb{R}^m$. The attention $a(\phi_x, \phi_{x_i}) $ measures the similarity between the input $x$ and $x_i $ in the support set through the neural embedding $\phi$. 

In this study, we employ the following attention mechanism resembling a kernel ridge regressor:
\begin{eqnarray}
y = \vb{g}(\phi_x)^{\top} (G_{\phi} + \lambda I )^{-1} \vb{y}
\label{eq:MNN}
\end{eqnarray}
where $\vb{y}^{\top} = (1, y_1, \ldots, y_m) \in \mathbb{R}^{m+1}$, $\vb{g}(\phi_x)^{\top} = (1, k(\phi_x, \phi_{x_1}), \ldots, k(\phi_x, \phi_{x_m})) \in \mathbb{R}^{m+1}$, and $G_{\phi}$ is the $(m+1) \times (m + 1)$ Gram matrix of positive definite kernels $k(\phi_{x_i}, \phi_{x_j})$ defined as
\begin{eqnarray*}
G_{\phi} &=&
\begin{bmatrix}
1       & 1                         & \dots     & 1 \\
1       & k(\phi_{x_1}, \phi_{x_1}) & \dots     & k(\phi_{x_1}, \phi_{x_m}) \\
\vdots  & \vdots                    & \ddots    & \vdots \\
1       & k(\phi_{x_m}, \phi_{x_1}) & \dots     & k(\phi_{x_m}, \phi_{x_m})
\end{bmatrix} 
\end{eqnarray*}
In Eq. \ref{eq:MNN}, $I$ is the $(m+1) \times (m+1)$ identity matrix, and $\lambda \in \mathbb{R}$ represents a controllable smoothing parameter. Note that element 1 is included in $\vb{y}$, $\vb{g}(\phi_x)$, and $G_{\phi}$ to introduce an intercept term into the regressor.
Here, $\vb{a}(\phi_x)^{\top} = \vb{g}(\phi_x)^{\top} (G_{\phi} + \lambda I )^{-1} $ in relation to Eq. \ref{eq:attention}.  In \citet{Bertinetto2019-ey}, this model was proposed as a differentiable closed-form solver in the context of few-shot learning using the model-agnostic meta-learning (MAML) \citep{Finn2017-ae} to obtain a meta-learner rapidly adaptable to a variety of tasks.

The E2T learning is formulated as the $\ell_2$ loss minimization:
\begin{eqnarray}
J_{\phi} &=& \sum_{(x_i, y_i, \mathcal{S}_i) \in \mathcal{T}} \Big(y_i - f(x_i, \mathcal{S}_i) \Big)^2 \nonumber \\
&=& \sum_{i=1}^n \sum_{(x_j^{\prime}, y_j^{\prime}) \in \mathcal{S}_i} \Big(y_i - a(\phi_{x_i}, \phi_{x_j^{\prime}}) y_j^{\prime} \Big)^2
\end{eqnarray}
In the two case studies presented below, we model the feature embedding $\phi$ by neural networks (see the Methods section for details).

The method of generating episodes involves different strategies in each case study. Intuitively, it is natural to include both extrapolative and interpolative episodes into a dataset, rather than solely relying on extrapolative episodes. The mixing rate of extrapolative to interpolative episodes would influence learning performance. It is also important to see that the size of $\mathcal{S}$ can be adjusted arbitrarily. In particular, the size of $\mathcal{S}$ can differ in the training and inference phases. Increasing $\mathcal{S}$ escalates the computational burden, particularly calculating the inverse matrix in Eq. \ref{eq:MNN}. To mitigate the computational cost, randomly sampled $\mathcal{S}$ should be used. Here, the question arises regarding the optimal size of $\mathcal{S}$ during the training and inference phases. 
To address these questions, we conducted various numerical experiments across the two case studies.

\subsection*{Experimental results}
The learning behavior and potential mechanisms of E2T were experimentally investigated in terms of predicting properties for materials out of the training sets. Here, we present performance evaluation experiments  focusing on extrapolative prediction tasks
for amorphous polymers and organic--inorganic hybrid perovskites.

\subsubsection*{Property prediction of out-of-domain polymers}
E2T was applied to a dataset of polymer properties calculated using RadonPy \citep{Hayashi2022-sw}. RadonPy is a software tool to automate the overall process of all-atom classical molecular dynamics (MD) simulations for various polymeric properties, including the specific heat at constant pressure ($C_p$) and refractive index. The dataset encompasses 69,480 amorphous polymers, which are classified into 20 polymer classes according to the chemical structures of their repeating units, such as polyimide, polyester, polystyrene, and so on (see Table S1 for the list of polymer classes and the number of polymers). The visualization of the chemical space using UMAP \citep{McInnes2018-yb} shows that these polymer classes are structurally distinct (Fig. S1), indicating that the prediction tasks across different polymer classes are extrapolative.

To evaluate the predictive performance regarding an unseen polymer class, the following procedure was conducted: (1) a model was trained using randomly chosen samples from 19 out of the 20 polymer classes, and (2) its generalization capability was accessed using data from the remaining polymer class. Two tasks were performed to predict $C_p$ and refractive index, respectively. The chemical structure of a polymer repeating unit was encoded with the Morgan fingerprint \citep{Morgan1965-mu, Rogers2010-ra} into a 2,048-dimensional descriptor vector, which serves as input for a three-layer fully connected neural network (FCNN) acting as the embedding function $\phi$ of MNN. As a baseline, a conventional FCNN with three hidden layers, which has an architecture similar to the embedding function of the MNN, was subjected to ordinary supervised learning. 

We assessed the scalability of the models' generalization capability on the sample size in the training dataset $\mathcal{D}$ and the support set $\mathcal{S}$, respectively. In each step of E2T, a training instance on $(x,y)$ was sampled from a randomly selected polymer class, while the support set $\mathcal{S}$ was sampled entirely from the 19 polymer classes including interpolative and extrapolative episodes. Throughout the training process, the size of the support set was fixed at $m=30$, whereas during the inference phase, the overall training dataset $\mathcal{D}$ was set to $\mathcal{S}$. The hyperparameter $\lambda$ was set to be 0.1, where its influence on the resulting out-of-domain generalization performance was investigated through the sensitivity analysis shown later. These experiments were repeated independently 10 times to calculate the mean predictive accuracy with their variability. Further details are described in the Methods section.

\begin{figure}[H]
    \centering
    \includegraphics[width=\textwidth]{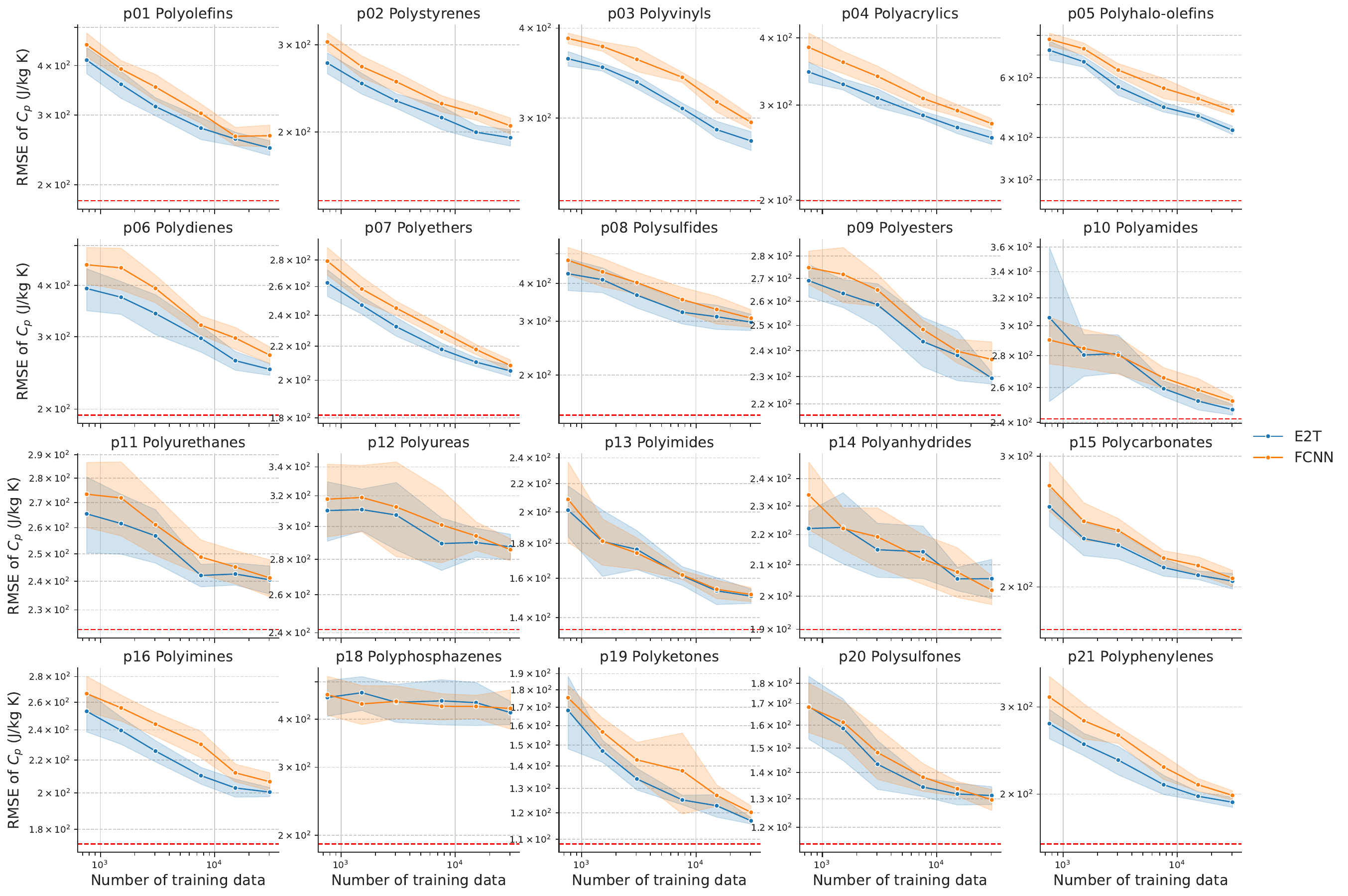}
    \caption{Scaling behavior of the out-of-domain generalization performance (RMSE: root mean squared error) of the specific heat ($C_p$) prediction task with the increasing number of training samples. RMSEs of MNNs trained with E2T and conventional FCNNs are shown in blue and orange, respectively. The red dashed lines denote the generalization performance of conventional domain-inclusive learning using data from all polymer classes.}\label{fig:radonpy_cp_rmse}
\end{figure}

\begin{figure}[H]
    \centering
    \includegraphics[width=\textwidth]{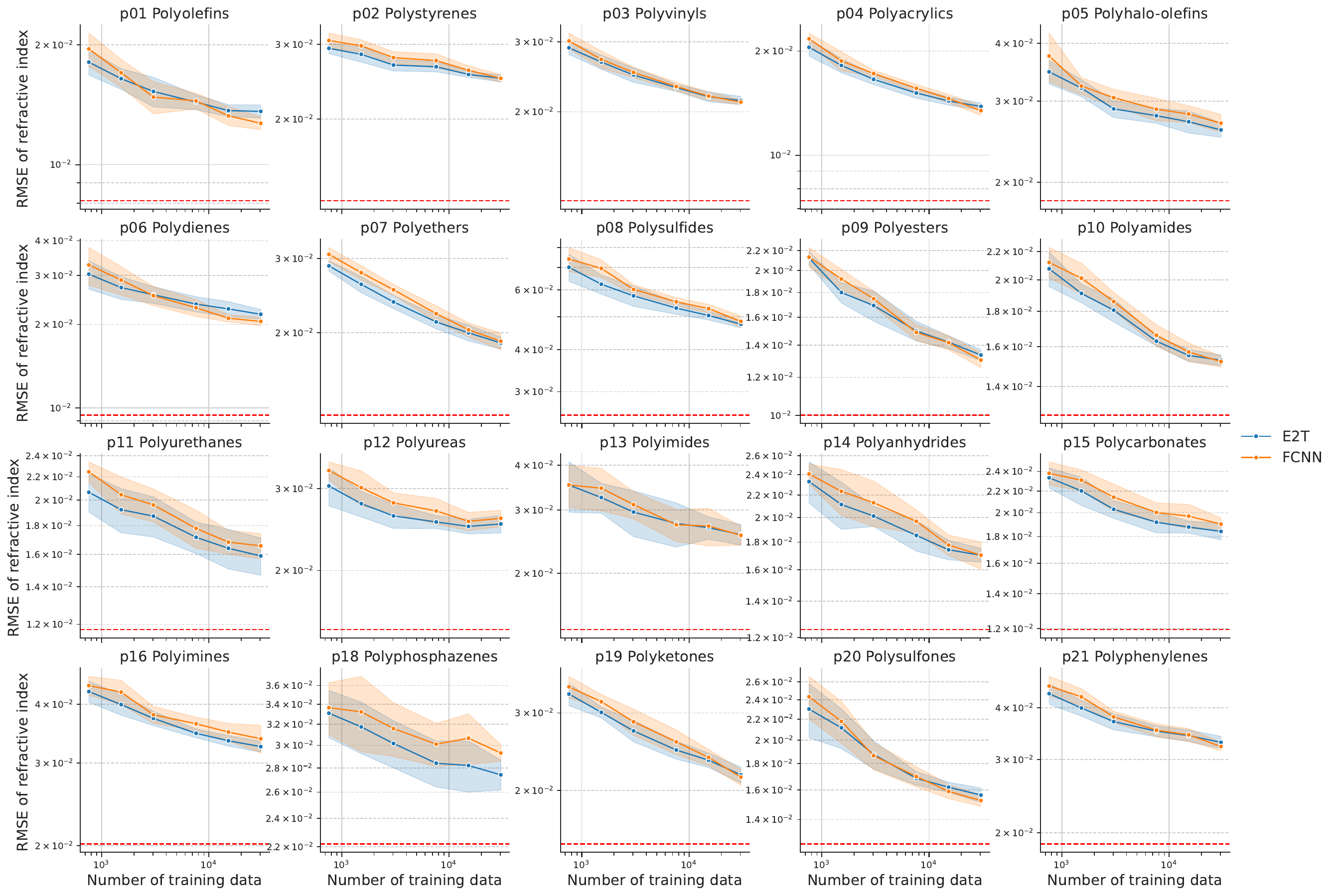}
    \caption{Scaling behavior of the out-of-domain generalization performance (RMSE) of the refractive index prediction task with the increasing number of training samples. RMSEs of MNNs trained with E2T and conventional FCNNs are shown in blue and orange, respectively. The red dashed lines denote the generalization performance of conventional domain-inclusive learning using data from all polymer classes.}\label{fig:radonpy_ri_rmse}
\end{figure}

Figs. \ref{fig:radonpy_cp_rmse} and \ref{fig:radonpy_ri_rmse} summarize the out-of-domain predictive performance on each of the 20 unseen polymer classes, improving almost monotonically to the increasing size of the training set $\mathcal{D}$. In each task on $C_p$ (Fig. \ref{fig:radonpy_cp_rmse}) or refractive index (Fig. \ref{fig:radonpy_ri_rmse}), E2T consistently and significantly outperformed the ordinary supervised learning with FCNN for most polymer classes across the different size of $\mathcal{D}$ varying in the range $[950, 38000]$, respectively. The generalization capability of E2T was scaled according to a power law with increasing training set on approximately the same order of magnitude as the ordinary supervised learning. In particular, there were no cases where E2T significantly underperformed compared to the ordinary learning. There were several polymer classes such as polyimides (p13), polyanhydrides (p14), and polyphosphazenes (p18) in the prediction of $C_p$, where E2T does not show notable improvement. Unfortunately, the underlying cause for the lack of improvement in several polymer classes could not be identified. The distributional features of property values for each polymer class, as shown in Fig. S2, did not exhibit any notable pattern associated with the observed extrapolative behaviors. In addition, the structure visualization using the UMAP projections in Fig. S1 did not reveal any structural uniqueness of these unsuccessful polymer classes. For instance, while p14 and p18 exhibited no significant improvement in the $C_p$ prediction task. E2T displayed substantial improvement over the ordinary learning in the refractive index prediction. This observation indicates that the potential gain in extrapolative prediction does not stem from cross-domain structural relationships but rather from the potential transferability regarding the presence or absence of physicochemical mechanisms.

In addition, we investigated the generalization performance of the FCNN trained on approximately 55,580 samples randomly chosen entirely from all polymer classes containing the target domain. As shown in Fig. \ref{fig:radonpy_cp_rmse} and Fig. \ref{fig:radonpy_ri_rmse} with red dashed lines, for many of the polymer classes, the extrapolative capability of E2T could not reach the level of the interpolative prediction of this baseline model. This suggests that while E2T enhances the extrapolative performance, it does not gain fundamental extrapolation capability. However, as demonstrated later, E2T can attain generalization performance equal to or significantly better than the baseline with much fewer training samples when fine-tuned to the target domain. This implies that extrapolatively trained models can adapt to a target domain rapidly with a small dataset.


\subsubsection*{Bandgap prediction of out-of-domain perovskite compounds}

To verify the generality of E2T, we conduct another experiment using the hybrid organic--inorganic perovskite (HOIP) dataset \citep{Kim2017-ga}. This dataset records 1,345 perovskite structures with their properties, including bandgaps, calculated by density functional theory. Each perovskite consists of a combination of organic/inorganic cations and an inorganic anion. The inorganic elements in the cations consist of germanium (Ge), tin (Sn), and lead (Pb), and the anions consist of fluorine (F), chlorine (Cl), bromine (Br), and iodine (I). Na et al. \citep{Na2022-rx, Na2022-uk} demonstrated state-of-the-art extrapolation performance using the automated nonlinearity encoder (ANE) on the HOIP dataset. ANE aims to enhance extrapolative prediction capability by utilizing an embedding function of input crystal structures that is pre-trained through self-supervised learning based on deep metric learning. Specifically, the embedding function was trained by minimizing the Wasserstein distance between the distances of given data in the embedding and property spaces, followed by ordinary supervised learning to predict physical properties using the embedded crystal structures. They considered two tasks mimicking real-world scenarios in exploring novel solar materials: (1) excluding perovskites containing both Ge and F from the training dataset, and (2) excluding perovskites containing both Pb and I from the training dataset. We refer to these tasks as ``HOIP-GeF'' and ``HOIP-PbI'', respectively. As shown in Fig. S3, in both tasks, the distributions of the training and test sets are extrapolatively related in both the structure and property spaces. In particular, the bandgap distribution of HOIP-GeF or HOIP-PbI is significantly biased toward the higher or lower tail respectively. 

We performed numerical experiments in the same setting as Na et al. \citep{Na2022-rx, Na2022-uk}. We divided the HOIP dataset into twelve groups based on combinations of four anions and three cations. When creating a training episode, we excluded the HOIP-GeF or HOIP-PbI dataset, and randomly selected 50 instances of $(x, y)$ from one group, while drawing $\mathcal{S}$ of size 50 from the remaining groups. In total, 1,248 or 1,228 training samples were drawn from the overall data other than HOIP-GeF or HOIP-PbI, respectively. Further experimental details are given in the Methods section. In the inference phase, the entire training dataset $\mathcal{D}$ was given to $\mathcal{S}$.

We examined the performance of E2T in comparison with ANE and conventional supervised learning. ANE and E2T were modeled by an embedding function followed by a regression header responsible for computing the bandgap as output. For the embedding of input crystal structures, the message passing neural network (MPNN) \citep{Gilmer2017-xa}, a graph neural network, was used for ANE and E2T. As the models for the header part, ANE and E2T employed FCNN and the kernel ridge regressor, respectively. As for the additional baselines, we used ``MPNN-Linear'' with the linearly modeled top layer and ``MPNN-FCNN''. 


The assessment of out-of-domain prediction accuracy is summarized in Table \ref{tab:hoip}. Similar to the polymer property prediction tasks, E2T showcased extrapolation performance that overwhelmingly surpassed the conventional learning (MPNN-Linear and MPNN-FCNN) for both tasks of HOIP-GeF and HOIP-PbI. Moreover, the extrapolation capability of E2T significantly exceeded that of ANE. 
Interestingly, while E2T did not attain the baseline prediction performance of an ordinary FCNN trained using the entire dataset, including instances from the target domain, it achieved a performance level remarkably close to it (Table \ref{tab:hoip}). For instance, the coefficients of determination ($\rm{R^2}$) for HOIP-PbI were $ 0.605 \pm 0.057$ for E2T and $ 0.675 \pm 0.162$ for the baseline, respectively, while $\rm{R^2}$ of ANE was $ 0.510 \pm 0.108$; similar results were observed for HOIP-GeF. This suggests that E2T has indeed acquired an extrapolation mechanism.

In the episodic training framework, several hyperparameters, such as the size of the support set, need to be adjusted. We conducted an ablation study using the HOIP dataset to investigate the influence of the training and inference support sizes, $|\mathcal{S}_{\mathrm{train}}|$ and $|\mathcal{S}_{\mathrm{infer}}|$, and the smoothing parameter $\lambda$ for the ridge regressor head on the E2T performance. 

\begin{figure}[H]%
    \centering
    \includegraphics[width=\textwidth]{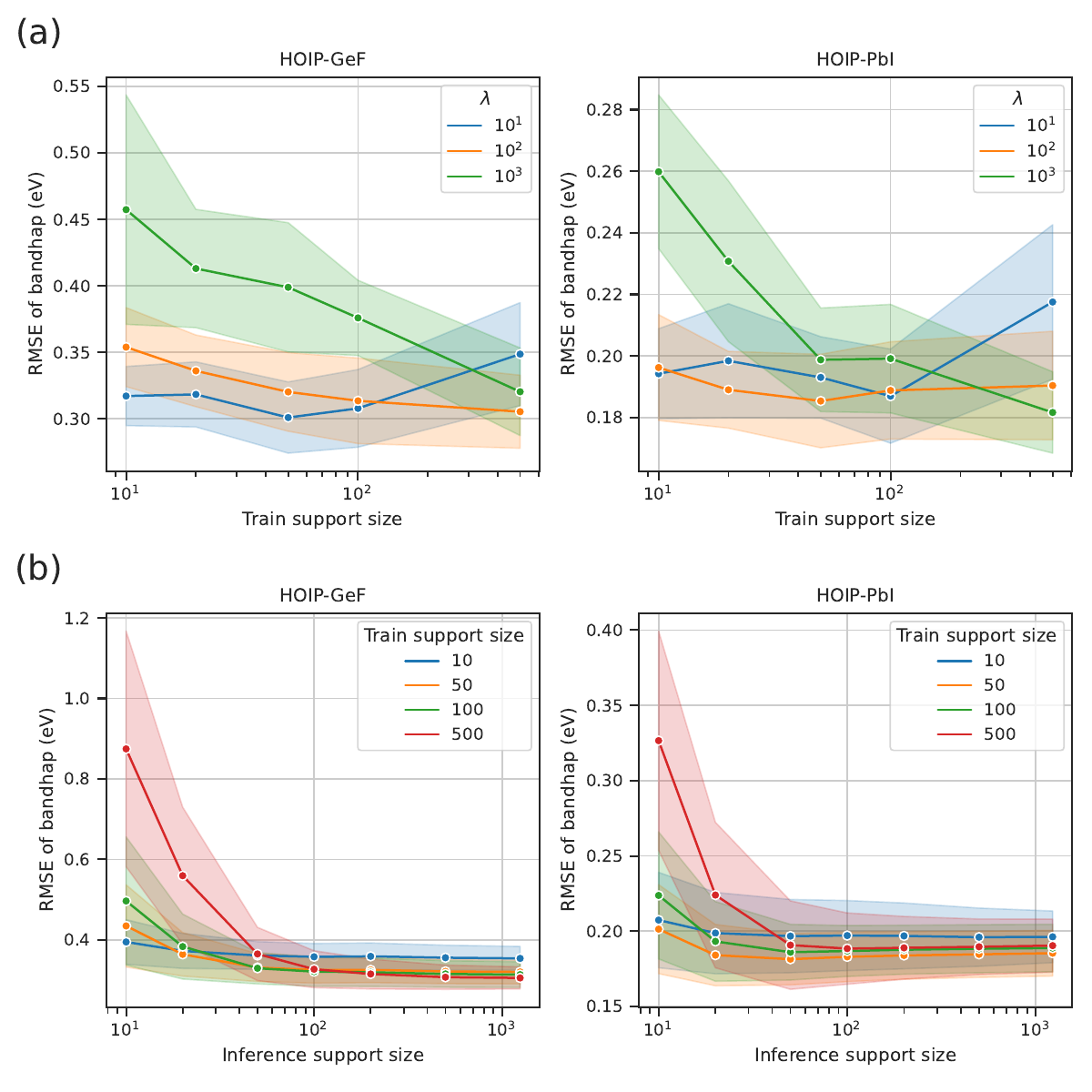}
    \caption{Sensitivity analysis of E2T in the two extrapolative prediction tasks (HOIP-GeF and HOIP-PbI) using the HOIP dataset. (a) Variation of the RMSE to varying the training support size with the inference support size fixed at 1,248 (left panel) and 1,228 (right panel). (b) Variation of the RMSE for varying the inference support size at $\lambda = 100$. In the panel (a), the colored lines indicate different smoothing parameters $\lambda$. In the panel (b), the colored lines represent the different training support sizes. The shaded areas indicate the standard deviations.
    }
    \label{fig:HOIP_ablation}
\end{figure}

As shown in Fig. \ref{fig:HOIP_ablation}(a), the generalization performance tends to improve with an increase in the training support size $|\mathcal{S}_{\mathrm{train}}|$, but the scaling behaviors were observed unclearly. Particularly in the HOIP-GeF task with the optimal support size $\lambda = 10$ exhibiting the best performance among the trials, the generalization performance did not change monotonically with increasing $|\mathcal{S}_{\mathrm{train}}|$. In summary, it is practically appropriate to keep the training support set relatively small, while controlling the value of $\lambda$ appropriately. 

Conversely, as shown in Fig. \ref{fig:HOIP_ablation}(b), the generalization performance scales monotonically with an increase in the inference support size $|\mathcal{S}_{\mathrm{infer}}|$. However, the decay of generalization performance nearly halts around $|\mathcal{S}_{\mathrm{infer}}| \approx 10^2$, regardless of the size of the training support set. From this test, it is concluded that setting $|\mathcal{S}_{\mathrm{infer}}| \approx 10^3$ is adequate to achieve satisfactory accuracy. 
In summary, it is preferable to use a large support set for inference, while ensuring an appropriate value of $\lambda$. In practice, $|\mathcal{S}_{\mathrm{infer}}|$ should be taken to be sufficiently large relative to $|\mathcal{S}_{\mathrm{train}}|$ under the constraint of computational cost.

\begin{table}
    \centering
    \caption{Evaluation of extrapolation prediction performance (RMSE and $\mathrm{R^2}$) based on the HOIP dataset. Two benchmark sets (HOIP-GeF and HOIP-PbI), excluding perovskite compounds with specific constituent elements, were used to predict the band gap of the unseen extrapolative compounds. MPNN-Linear (all) refers to non-extrapolative models trained using data from the entire domain including the target. We conducted 30 runs independently and the standard deviation of the performance metrics is indicated after the symbol $\pm$.}
    \begin{tabular}{ccccc}
        \toprule
        \multirow{2}{*}{\text{Methods}} & \multicolumn{2}{c}{\text{HOIP-GeF}} & \multicolumn{2}{c}{\text{HOIP-PbI}} \\
        \cmidrule(lr){2-3} \cmidrule(lr){4-5}
        & $\rm{R^2}$ & $\rm{RMSE}$ (eV) & $\rm{R^2}$ & $\rm{RMSE}$ (eV) \\
        \midrule
        MPNN-Linear & $0.255 \pm 0.198$& $0.361 \pm 0.046$ & $0.545 \pm 0.064$ & $0.207 \pm 0.014$ \\
        MPNN-FCNN & $-0.088 \pm 0.614$& $0.427 \pm 0.106$ & $0.508 \pm 0.185$ & $0.213 \pm 0.037$ \\
        ANE \cite{Na2022-rx} & $0.361 \pm 0.105$ & $0.336 \pm 0.027$ & $0.510 \pm 0.108$ & $0.214 \pm 0.024$ \\
        E2T & $\mathbf{0.486 \pm 0.095}$ & $\mathbf{0.301 \pm 0.027}$ & $\mathbf{0.605 \pm 0.057}$ & $\mathbf{0.193 \pm 0.013}$ \\
        \midrule
        MPNN-Linear (all) & $0.551 \pm 0.418$ & $0.244 \pm 0.045$ & $0.675 \pm 0.162$ & $0.168 \pm 0.037$ \\
        \bottomrule
    \end{tabular}
    \label{tab:hoip}
\end{table}

\subsection*{Fine-tuning to extrapolative domains}

So far, our focus has been on scenarios where no data are available during the episodic training for the target domain. Here, we shift our attention to scenarios where a limited amount of data is accessible in the target domain such scenarios are common in practical materials development. In such cases, leveraging data from a related source domain via transfer learning including fine-tuning is a pragmatic approach \citep{Yamada2019-sz, Wu2019-oz}. Moreover, meta-learning methods have proven effective for few-shot classification problems, such as toxicity prediction \citep{Altae-Tran2017-bo, Ju2023-uv, Vella2023-zu}. Inspired by these previous studies, we adapted a pre-trained meta-learner to data from the target domain via fine-tuning, as detailed in the Methods section. Below, we present the results of applying our methodology to the two distinct problem settings.

The fine-tuning was performed on the RadonPy dataset. In this experiment, a pre-trained model of E2T with a source data size of 38,000 was fine-tuned using data from the target domain corresponding to a particular polymer class. To fine-tune a pre-trained model of E2T, episodes $(x_i, y_i, \mathcal{S}_i)$ were randomly sampled from all data containing the polymer class of the target domain to modify the entire network. The pre-trained FCNNs underwent fine-tuning across all layers of their respective networks using data from the target polymer class. 

\begin{figure}[H]
    \centering
    \includegraphics[width=\textwidth]{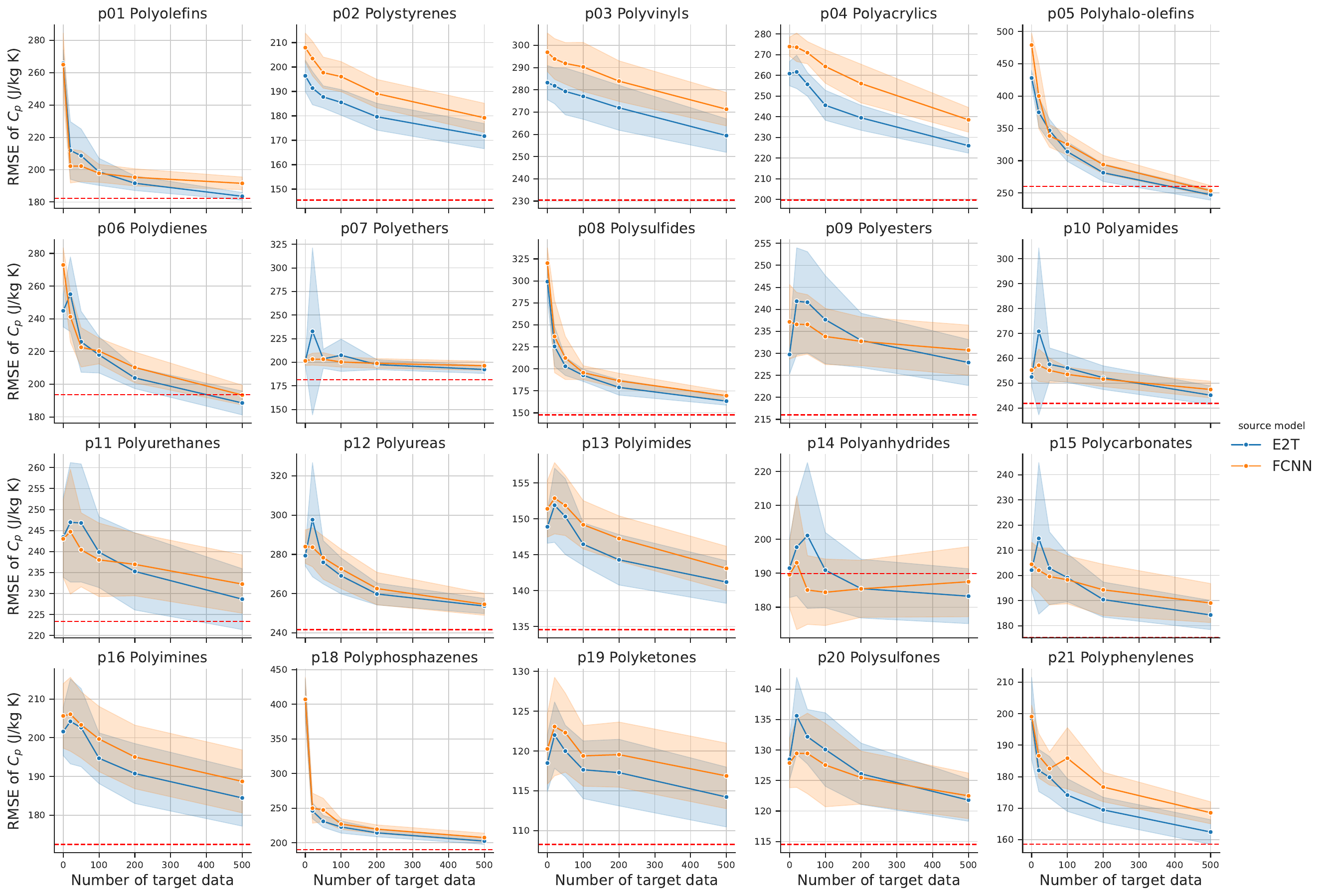}
    \caption{
    Scaling behavior of the fine-tuned $C_p$ predictor with increasing target samples. The results of E2T are depicted in blue, while FCNN is shown in orange. Each panel represents a different polymer class. The x-axis indicates the number of samples from the target domain for fine-tuning, while the y-axis represents RMSE with the standard deviation. The red dashed line denotes the generalization performance of the model trained on the entirely sampled dataset, including the target domain.}\label{fig:radonpy_ft_cp_rmse_s38000}
\end{figure}

\begin{figure}[H]
    \centering
    \includegraphics[width=\textwidth]{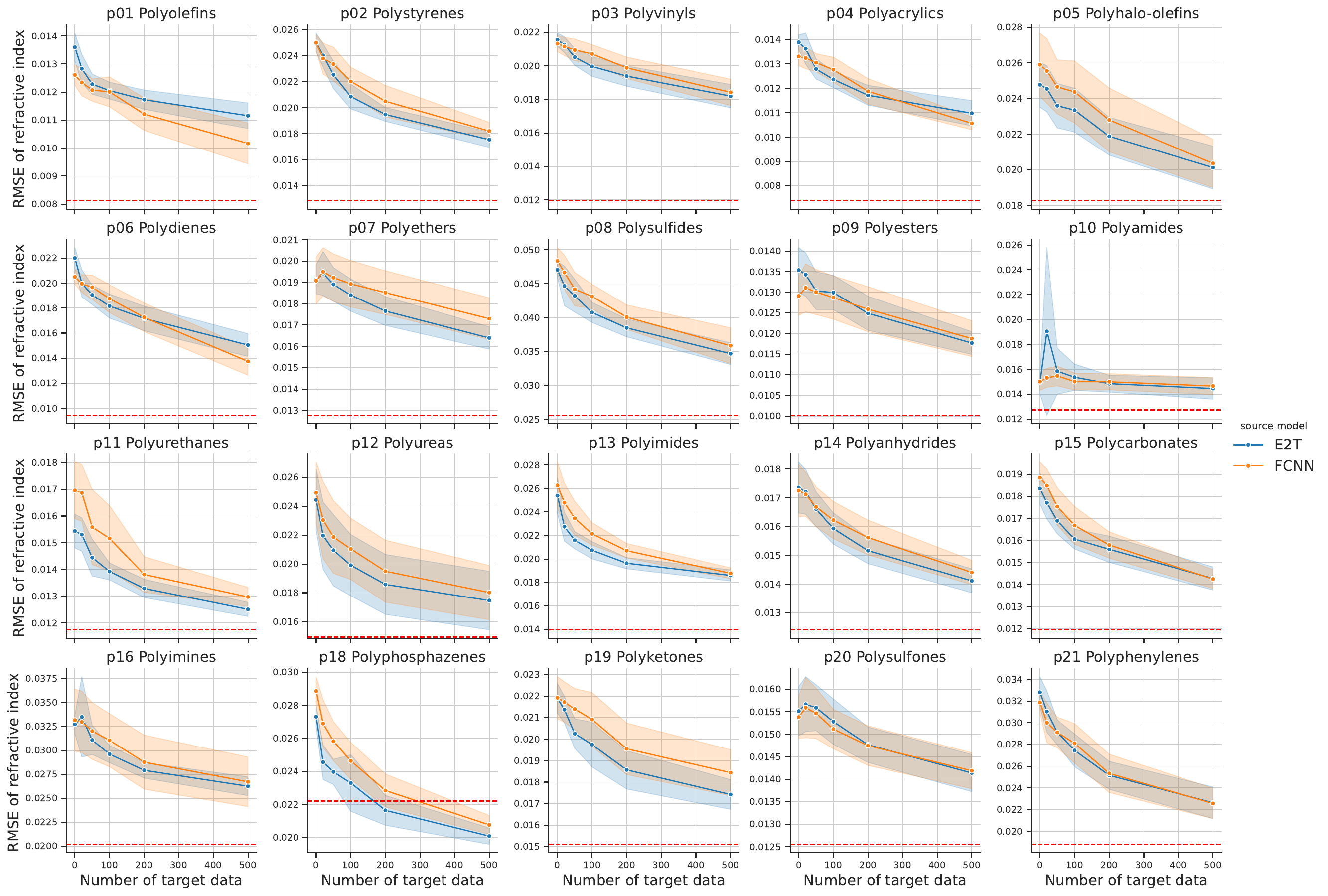}
    \caption{
    Scaling behavior of the fine-tuned refractive index predictor with increasing target samples. The results of E2T are depicted in blue, while FCNN is shown in orange. Each panel represents a different polymer class. The x-axis indicates the number of samples from the target domain for fine-tuning, while the y-axis represents RMSE with the standard deviation. The red dashed line denotes the generalization performance of the model trained on the entirely sampled dataset, including the target domain.}\label{fig:radonpy_ft_ri_rmse_s38000}
\end{figure}

As shown in Figs. \ref{fig:radonpy_ft_cp_rmse_s38000} and \ref{fig:radonpy_ft_ri_rmse_s38000}, the loss decreases as the target data increases almost monotonically for E2T and FCNN. Focusing on the difference between E2T and FCNN, E2T outperforms FCNN in most of the cases, implying the superiority of E2T over ordinary supervised learning even in fine-tuning scenarios. In particular, E2T scaled with no order-level differences, but maintained gains constantly for increasing numbers of trained data. Furthermore, as before, comparisons were also made with the baseline FCNNs trained on the entire dataset, including samples from the target domain.

Notably, for example, the $C_p$ prediction performance of E2T in polyhalo-olefins (p05) and polydienes (p06) reached the baseline performance indicated by the red dashed lines in the figure. For training the baseline model, 1,154 and 1,047 samples were used for p05 and p06, respectively. In contrast, only 500 or fewer samples were used to fine-tune the MNNs to achieve the same level of performance. For the other polymer classes, according to their scaling behaviors, it is estimated that the baseline performance will be exceeded by the one of E2T with considerably fewer samples, suggesting that models extrapolatively trained by E2T can adapt early to inexperienced domains.


\begin{figure}[htbp]
    \centering
    \includegraphics[width=\textwidth]{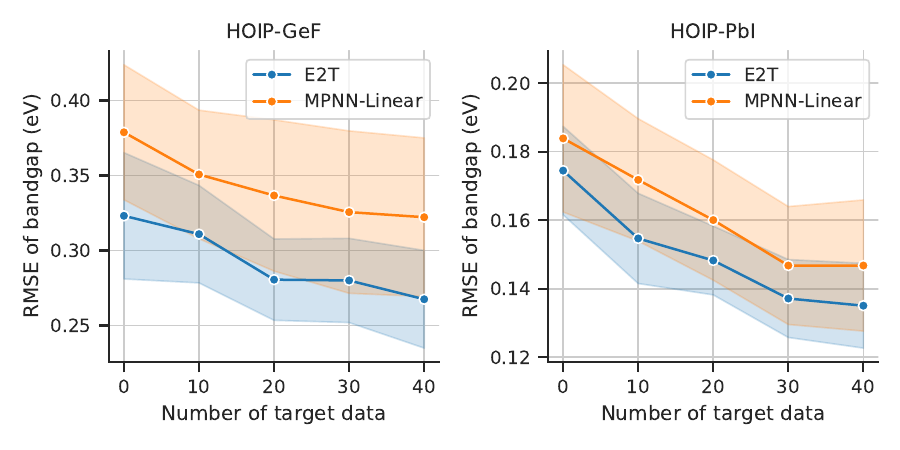}
    \caption{Scaling behavior of bandgap prediction loss as the number of target samples increases. The left and right panels represent the results for HOIP-GeF and HOIP-PbI, respectively. The results of E2T and MPNN-Linear are distinguished by blue and orange colors, respectively. The x-axis denotes the number of target samples used for fine-tuning, while the y-axis denotes the RMSE of the bandgap predictions along with the standard deviation.}
    \label{fig:hoip_ft_rmse}
\end{figure}

Similar experiments were conducted on the HOIP dataset, where MNNs pre-trained with E2T on 1,248 or 1,228 source datasets were transferred to predict for the target domains, namely HOIP-GeF and HOIP-PbI. Episodes for fine-tuning were generated using samples from the source and target domains. In contrast, the pre-trained MPNN-Linear model was fine-tuned solely using data from the target domain. The scaling behaviors are illustrated in Fig. \ref{fig:hoip_ft_rmse}, highlighting that E2T outperforms the ordinary supervised learning, thereby supporting the conclusions based on the experimental results obtained using the RadonPy dataset.



\section*{Discussion}\label{discussion}
Predicting material properties beyond the range of data distribution is the ultimate goal of materials science. This study has presented a machine learning methodology to address this fundamental challenge. Previous approaches have relied on incorporating physical prior knowledge into models as descriptors or by adding known theories or empirical rules to model architectures through methods like physics-informed machine learning, aiming to extract extrapolative predictability. In contrast, we set out to achieve extrapolation capability through fully inductive reasoning without any physical insights. Specifically, we focused on the MNN architecture proposed in few-shot learning and used it as a meta-learner to solve extrapolative prediction tasks. It was demonstrated that the meta-learner could indeed acquire outstanding out-of-domain generalization capability through experiencing numerous extrapolative tasks generated by the E2T algorithm. In this study, while the generalization performance of the meta-learner did not reach the achievable limit of an oracle model learned from all datasets including the target extrapolation domain, the significance of the improvement in extrapolation performance compared to the baseline was substantial in most cases. Furthermore, it was experimentally confirmed that meta-learners trained with extrapolative training could quickly transfer to unexplored domains with a small amount of additional data, suggesting the early adaptation capability of learners trained to tackle challenging problems.

Our study is still in the first step, and several technical challenges and research questions remain to be addressed. Computing MNNs requires keeping past training data in memory as the support set, which has its limitations in terms of the data volume that can be retained. Additionally, as the data volume increases, the computational load of the kernel ridge regression header also increases. The retention of data in memory also raises privacy concerns. Leveraging other methodologies of meta-learning such as MAML or its derivatives, could serve as a solution to these issues. When designing the method of generating episode sets, there are various hyperparameters to consider. In particular, the mixing ratio of interpolative and extrapolative episodes in the episode set is expected to impact generalization performance. For instance, a learner trained heavily on extrapolative episodes may not predict interpolative tasks appropriately. Generally, experiencing tasks of varying difficulty levels evenly is considered an appropriate learning method. Furthermore, it is also intriguing to investigate whether the observed early adaptability of meta-learners to new tasks holds universally.


\section*{Methods}\label{methods}

\subsection*{Polymer property prediction}
\subsubsection*{Data}
In the polymer property prediction experiments, 69,480 samples of $C_p$ and 68,700 samples of refractive index were used for amorphous homopolymers. The data were generated using RadonPy \citep{Hayashi2022-sw}, which is a software for calculating various physical properties of polymers using all-atom molecular dynamics simulations. This dataset includes 1,078 samples already available in open source and newly generated by the RadonPy consortium. Approximately 70,000 hypothetical polymers were generated using an N-gram-based polymer structure generator \citep{Ikebata2017-ra} and classified into 20 polymer classes based on the rule by PolyInfo \citep{Otsuka2011-mv}. The list of the 20 polymer classes with their data size is shown in Table S1.

\subsubsection*{Descriptor}
The count-based Morgan fingerprint \citep{Morgan1965-mu}, a type of extended connectivity fingerprints (ECFP) \citep{Rogers2010-ra}, was utilized as a descriptor of the repeating unit of homopolymer. The descriptor calculation was performed using RDKit \citep{rdkit}, with the selected parameters being a radius of 3 and a bit length of 2,048. 

\subsubsection*{Training of MNNs by E2T}
The attention-based model resembling a kernel ridge regressor of Eq. \ref{eq:MNN} was implemented in PyTorch \citep{Paszke2019-lb}. The three-layer fully connected neural network with ReLU activation was used as an embedding function $\phi$ from the 2,048-dimensional descriptor to the 16-dimensional latent space. The layer structure of $\phi$ was configured with neurons of 2048, 128, 128, and 16, respectively, and the last 16-dimensional vector was normalized using layer normalization \citep{Ba2016-wo}. As for the ridge regressor head, a smoothing parameter was set at $\lambda=0.1$.

The data from 19 polymer classes out of 20 were used for training and testing to evaluate the out-of-domain prediction performance. To investigate the influence of the training data size on the generalization performances, the size of training samples was varied as $|\mathcal{D}| \in \{950, 1900, 3800, 9500, 19000, 38000\}$. The training set was generated from 19 polymer classes so that the number of samples from each class becomes the same. Each training set was further split into training $\mathcal{D}_\mathrm{train}$ and validation $\mathcal{D}_\mathrm{val}$ with the proportions of $80\,\%$ and $20\,\%$, respectively. In each step of E2T, a training instance on $(x, y)$ was sampled from a randomly selected polymer class, while the support set $\mathcal{S}$ of the size $m=30$ was sampled entirely from the 19 polymer classes including interpolative and extrapolative episodes. The prediction performance was monitored by loss
\begin{eqnarray*}
\frac{1}{|\mathcal{D}_\mathrm{val}|}\sum_{(x,y)\in \mathcal{D}_\mathrm{val}}\big(y-f_{\phi}(x, \mathcal{D}_\mathrm{train})\big)^2    
\end{eqnarray*}
The training was halted when observing no improvement over 90,000 episodes. The training was performed with a dropout rate \citep{Srivastava2014-ak} of 0.2 and a constant learning rate of $2\times10^{-4}$ with the Adam optimizer \citep{Kingma2014-dv}. The trained model was evaluated on the data from the remaining polymer class. These experiments were repeated 10 times for each condition with different random seeds.


\subsubsection*{Training of fully connected networks by ordinary supervised learning}
The four-layer fully connected neural networks configured with 2048, 128, 128, 16, and 1 neurons were implemented in PyTorch. ReLU was used for the activation function, and layer normalization was applied to the 16-dimensional hidden representation.
Data from 19 out of 20 polymer classes were used for training and the remaining class was used to evaluate the performance of out-of-domain prediction. To investigate the influence of the size of the training dataset on the performance, different sizes of dataset $|\mathcal{D}| \in \{950, 1900, 3800, 9500, 19000, 38000\}$ were sampled from the dataset in the 19 classes so that the number of samples from each polymer class is equal. $20\,\%$ of the training set was used for the validation set. The training was performed with a dropout rate of 0.2, a batch size of 256, and a constant learning rate of $2\times10^{-4}$ using the Adam optimizer. The training was terminated when no improvement was observed for 50 epochs. The trained model was evaluated on the data from the remaining polymer class. The experiment was repeated 10 times for each condition with different random seeds.

\subsubsection*{Generalization performance of domain-inclusive learning}
Four-layer neural networks were implemented to evaluate the generalization performances of domain-inclusive learning using data from the entire chemical space. The overall data including 20 polymer classes was split into training, validation, and test sets with the proportion $64\,\%$, $16\,\%$, and $20\,\%$ respectively. Using the training and validation sets, the network was trained using the same procedure as the training of the fully connected models in the out-of-domain task. The trained model was evaluated on data from the test set for each polymer class. The experiment was repeated 5 times with different data splits.

\subsection*{Bandgap prediction of perovskite compounds}
\subsubsection*{Data}
The hybrid organic--inorganic perovskites (HOIP) dataset was used in this experiment \citep{Kim2017-ga}. This dataset contains 1,345 perovskite compounds with their properties, including bandgap, dielectric constant, and relative energies, calculated by density functional theory. Each compound consists of an organic cation, an inorganic cation, and an inorganic anion. The inorganic elements consist of Ge, Sn, and Pb cations and F, Cl, Br, and I anions. In this experiment, as in prior works \citep{Na2022-rx, Na2022-uk}, we evaluated generalization performances in two different tasks: (1) bandgap prediction of perovskite compounds containing Ge and F, and (2) prediction of perovskite compounds containing Pb and I. The former and latter sets exhibit extremely higher or lower bandgaps. 

\subsubsection*{Embedding function of crystal structures}
The message passing neural network (MPNN) \citep{Gilmer2017-xa} was employed as an encoder of crystal structures for all models. The MPNN architecture was designed similarly to that in the work by Na and Park \cite{Na2022-rx}. The embedding size was set to 32, and detailed settings can be found in their GitHub repository https://github.com/ngs00/ane.

\subsubsection*{Training of MNNs by E2T}
The attention with kernel ridge regressor of Eq. \ref{eq:MNN} was implemented in PyTorch. In the HOIP experiment, MPNN was employed as an embedding function $\phi$ that transforms an input crystal structure to the 32-dimensional latent vector. The embedding variable was normalized by performing layer normalization. A smoothing parameter of the ridge regressor head was set at $\lambda=10$.
We classified the HOIPs dataset into twelve categories (or domains) based on the combination of four inorganic anions and three cations. The data from eleven out of the twelve categories were used for the training dataset $\mathcal{D}$. To monitor the model performance, $10\,\%$ of $\mathcal{D}$ was allocated for validation. In each step of E2T, a training instance at $(x, y)$ was sampled from a randomly selected combination from the eleven anion--cation combinations, while the support set $\mathcal{S}$ with the size of $m=50$ was sampled from the remaining ten combinations of anion and cation, resulting in the inclusion of only extrapolative episodes. The prediction performance was monitored using a validation set, and the training was stopped when no improvement was observed after 150,000 episodes. The training was performed with a constant learning rate of $5\times10^{-4}$ with the Adam optimizer. The trained model was evaluated on the data from a remaining anion--cation combination, i.e., HOIP-GeF or HOIP-PbI. The experiment was repeated 30 times for each condition with different random seeds.

\subsubsection*{ANE-MPNN}
The automated nonlinearity encoder (ANE) \citep{Na2022-rx} is the state-of-the-art method for extrapolation tasks to our best knowledge, as verified on HOIP dataset in the previous work. The training of the ANE method involves two stages: (1) pre-training through metric learning to obtain a feature embedding and (2) supervised learning for training the header network that maps the embedded input to its output. The previous work demonstrated that ANE with the MPNN encoder (ANE-MPNN) outperforms several other models. We trained ANE-MPNN based on the settings described in the original paper and the distributed code. Specifically, the MPNN encoder was trained with a learning rate of $1\times10^{-3}$ and a batch size of 32. The header network, consisting of four layers of the size 32, 356, 128, and 1, was trained with a learning rate of $5\times10^{-4}$, an $\ell_2$ regularization coefficient of $1\times10^{-6}$, and a batch size of 64. We trained the models for 500 epochs without early stopping. The experiment was conducted 30 times with different random seeds.

\subsubsection*{Baseline: MPNN-Linear and MPNN-FCNN}
As a baseline for conventional feed-forward supervised learning, we trained two models consisting of an MPNN encoder and an FCNN/Linear header. The first model, serving as a counterpart to E2T, used a single linear layer as a header and is referred to as MPNN-Linear. The other model, MPNN-FCNN, utilized an FCNN header with the same architecture as ANE-MPNN. Layer normalization was applied to the embedding vector produced by the MPNN in both models. The data excluding compounds containing both Ge and F, or both Pb and I were used for the training dataset. To monitor the change in generalization performance during training, $10\,\%$ of the training dataset was allocated for validation. The training was performed with a batch size of 128, and a constant learning rate of $5\times10^{-4}$ using the Adam optimizer The training was terminated upon observing no improvement for 300 epochs. The experiment was conducted 30 times with different random seeds.

\subsubsection*{Generalization performance of domain-inclusive learning}
An architecture similar to the MPNN-Linear models was implemented to evaluate the prediction performance of domain-inclusive learning. The overall dataset was split into 72:8:20 for training, validation, and testing. Using the training and validation sets, the model was trained by performing the same procedure as the out-of-domain prediction tasks. The trained model was evaluated on the HOIP-GeF or HOIP-PbI compounds with unseen chemical elements. The experiment was conducted 30 times with different data split patterns.

\subsubsection*{Sensitivity analysis of hyperparameters in E2T}
The extrapolative performance was evaluated by varying three hyperparameters: $\lambda$, $|\mathcal{S}_\mathrm{train}|$, and $|\mathcal{S}_\mathrm{infer}|$. The model was trained 30 times with different random seeds for each pair of $\lambda \in \{10, 100, 1000\}$ and $|\mathcal{S}_\mathrm{train}| \in \{10, 20, 50, 100, 500\}$. The extrapolative prediction of each trained model was performed with different sizes of inference support set $|\mathcal{S}_\mathrm{infer}| \in \{10, 20, 50, 100, 500, 1248\}$ or $\{10, 20, 50, 100, 500, 1228\}$ for HOIP-GeF and HOIP-PbI, respectively. The support $|\mathcal{S}_\mathrm{infer}|$ was sampled 10 times independently.

\subsection*{Fine-tuning experiments}
\subsubsection*{Polymer property prediction}
An MNN trained by E2T with a source data size of 38,000 was fine-tuned with data including samples in the target domain. Half of the target data was reserved for the performance evaluation, while 20 to 500 samples of the remaining data --specifically 20, 50, 100, 200, and 500 samples-- were used for fine-tuning. Episodes $(x_i, y_i, \mathcal{S}_i)$ were sampled from the source and target datasets to modify the pre-trained embedding function $\phi$. To monitor the model performance during the fine-tuning, 20 \% of the target dataset was allocated for validation and the training was stopped on observing no improvement over 60,000 episodes. The learning rate was set at $10^{-5}$. The size of the training support set was fixed at $m=20$. The experiment was conducted across all combinations of five different source models independently pre-trained on $\mathcal{D}$ with $d=38,000$ and nine different data splits, resulting in 45 runs for each polymer class and fine-tuning data size.

As a baseline in the comparative study, a fully connected neural network trained by ordinary supervised learning with a source data size of 38,000 was fine-tuned using data from the target domain. Half of the target dataset was set aside for evaluation, with sample sizes ranging from 20 to 500 from the remaining data used for fine-tuning. $20\,\%$ of the fine-tuning data was allocated for validation, and the training was stopped on observing no improvement over 50 epochs. The learning rate was set at $10^{-5}$. The batch size was set to one for fine-tuning with training data sized at 20 and 50 samples, while a batch size of 32 was used for the larger fine-tuning datasets. The experiment was executed across five independently obtained models and nine different data splits, resulting in 45 runs for each polymer class and dataset size.

\subsubsection*{Bandgap prediction of perovskite compounds}
An MNN trained by E2T with a source data size of 1,248 or 1,228 was fine-tuned using data including data from the target domain. Half of the target dataset was reserved for performance evaluation, while 10 to 40 samples from the remaining data were used for fine-tuning. Episodes $(x_i, y_i, \mathcal{S}_i)$ were sampled from the source and target datasets to refine the embedding function $\phi$. The model was trained over 3,000 episodes with a learning rate of $10^{-5}$. Early stopping was not applied for this experiment because the target data size was small. The size of the training support set was fixed at $m=10$. The experiment was conducted for each combination of 10 independently obtained models and four different data splits, resulting in a total of 40 runs for each data size.

A model of MPNN-Linear pre-trained by ordinary supervised learning with a source data size of 1,248 or 1,228 was fine-tuned using data from the target domain. Half of the target dataset was set aside for performance evaluation, and a subset of 10 to 40 samples from the remaining data was used for fine-tuning. The models underwent fine-tuning over 300 epochs, with a learning rate of $10^{-5}$ and a batch size of 10. The experiment was executed across 10 different models and four different data splits.

\bibliography{sn-bibliography}

\end{document}